\documentclass[11pt]{article}
\usepackage{moriond,epsfig}

\def\be{\begin{equation}}
\def\ee{\end{equation}}
\def\bea{\begin{eqnarray}}
\def\eea{\end{eqnarray}}

\begin{document}
\thispagestyle{empty}
\vspace*{4cm}
\title{CHEMICAL EVOLUTION OF CLUSTERS OF GALAXIES}
\author{ L.~PORTINARI }

\address{Theoretical Astrophysics Center, Juliane Maries Vej 30,\\
DK-2100 Copenhagen \O, Denmark}

\author{ A.~MORETTI and C.~CHIOSI }

\address{Dipartimento di Astronomia, Vicolo dell'Osservatorio 2,\\
I-35122 Padova, Italy}

\maketitle\abstracts{
The high metallicity of the intra--cluster medium (ICM) is 
generally interpreted on the base of the galactic wind scenario for 
elliptical galaxies. In this framework, we develop a toy--model to follow 
the chemical evolution of the ICM, formulated in
analogy to chemical models for individual galaxies. Just as the ingredients 
for usual models are 
(a) the stellar yields, amount of metals newly synthesized and re-ejected 
by stars;
(b) the Star Formation Rate and 
(c) the stellar Initial Mass Function (IMF), 
our model for clusters involves:
(a') ``galactic yields'' derived from galactic wind models of ellipticals;
(b') a parametric Galactic Formation Rate;
(c') a Press-Schechter-like Galactic Initial Mass Function.\\
The model is used to test the response of the predicted 
metal content and abundance evolution of the ICM to varying input
galactic models. 
The resulting luminosity function of cluster galaxies is also calculated,
in order to constrain model parameters.}
\section*{Introduction}
The popular galactic wind (GW) scenario, introduced by Larson (1974)
to account for the photometric properties of elliptical galaxies, 
predicted as a side effect the pollution of the intra--cluster medium 
(ICM) with the chemical elements produced and expelled by individual
galaxies (Larson \& Dinerstein 1975). 
Metals in the hot ICM were in fact detected soon afterwards (Mitchell 
et~al.\ 1976, Serlemitsos et~al.\ 1977).

The amount of metals present in the ICM, as reported by Renzini (1997), 
is currently estimated as:
\[ \frac{M_{ICM}^{Fe}}{M_*^{Fe}} = 
\frac{Z_{ICM}^{Fe} \, M_{ICM}}{Z_*^{Fe} \, M_*} = 1.65 \, h^{-3/2} \]
with obvious meaning of the symbols. Iron is generally used as tracer of the
overall metallicity, being the best measured element in the hot ICM.
With $h \sim 0.65$, three times more metals  seem to be spread in the ICM 
than locked into the stellar component of the individual galaxies.

The source of such a large amount of metals in the ICM could be galactic
(as in the original prediction by Larson) or reside in Population III
pre-galactic objects (White \& Rees 1978). The distinct correlation between
the iron mass in the ICM and the luminosity of elliptical and S0 galaxies,
\[ M_{ICM}^{Fe} \propto L_V^{E+S0}\]
demonstrated by Arnaud et~al.\ (1992), seems to favour 
galaxies as the sites of production of the metals in the ICM.

Accepting that the metals in the ICM originated in the E and S0 galaxies
of the cluster, two mechanisms exist to extract the newly
sinthesized elements from the individual galaxies: the above mentioned GW
and ram pressure stripping. Arguments have been given by Renzini (1997)
favouring the GW scenario, mostly based on the observation that the
``iron mass--to--luminosity ratio'' (IMLR) is roughly constant
independently of cluster richness and temperature, while the ram pressure 
mechanism should be more efficient, extracting more metals 
for a given stellar content, in richer clusters.

Though the role of ram pressure stripping is still debated 
(e.g.\ Mori \& Burkert 2000), from here on we will limit to the GW scenario
for the pollution of the ICM, bearing in mind that the addition of
other mechanisms (metal production in pre-galactic objects and/or 
ram pressure stripping) would allow to inject even more metals into the ICM,
further favouring the enrichment.
\section{The gas and metal content of the ICM}
\label{section1}
In this section we briefly summarize previous literature and current
understanding of the gas and metal content in the ICM, from the point
of view of theoretical models of galactic chemical evolution with GWs.
\subsection{The metal content of the ICM}
\label{metalcontent}
Various early studies investigated whether ``standard'' chemical models
for galaxies can explain the amount of metals detected in the ICM
(Vigroux 1977, Himmes \& Biermann 1980, De Young 1978); by ``standard''
we mean a chemical model with the same physical ingredients (mainly,
stellar Initial Mass Funstion and yields) suited
to reproduce the Solar Neighbourhood. Amidst these early studies, 
we mention in
particular the one by Matteucci \& Vettolani (1988), as the first
attempt to link directly the metallicity of the ICM with the properties
of the corresponding galaxy population. To this aim, the authors
developed a modelling technique that has been widely adopted afterwards.
Basing on a grid of models of elliptical galaxies 
with GW, they assigned to any given galaxy of final stellar mass $M_*$, 
or equivalently of final present--day luminosity $L_*$, the corresponding
masses of gas and iron ejected in the GW ($M^{ej}_{gas}$, $M^{ej}_{Fe}$).
Integrating these quantities over the observed luminosity function (LF), 
they calculated the total masses of gas and iron globally ejected
by the galactic population in the cluster. Their main conclusions
were:\\ 
(1) the iron content of the ICM can be reproduced with a standard 
Salpeter Initial Mass Function (IMF) in the individual galaxies;\\
(2) the global amount of gas ejected as GW is much smaller than the observed
mass of the ICM, hence the ICM must be mostly primordial gas which
was never involved in galaxy formation.

This early successful reproduction of $M_{Fe}^{ICM}$ turned out afterwards
to be favoured by
the low gravitational potential wells of model galaxies, calculated 
only on the base of their luminous, baryonic component. Once the potential
well of the much heavier dark matter halo is included, the ejecta of SN Ia
hardly escape the galaxy and the metal pollution of the ICM by GWs
is much reduced (David et~al.\ 1991, Matteucci \& Gibson 1995). In this case,
standard chemical models fail to reproduce the metal content of the ICM.
Some non--standard scenarios were thus invoked to solve the riddle, such as:
\begin{itemize}
\item
a more top--heavy IMF than the Salpeter one, with logarithmic slope 
$x \sim 1.0$ rather than the standard value $x=1.35$ 
(Matteucci \& Gibson 1995, Gibson \&
Matteucci 1997ab, Loewenstein \& Mushotzky 1996);
\item
a bimodal IMF with an early generation of massive stars heavily polluting
the ICM, followed by a more normal star formation phase
producing the stars we see today (Arnaud et~al.\ 1992, Elbaz et~al.\ 1995).
\end{itemize}
These models, where SN II from massive stars play the main role in the
metal pollution of the ICM, were further supported by the enhanced 
abundances  of $\alpha$--elements with respect to iron detected with ASCA 
(e.g.\ Mushotzky et~al.\ 1996; but see nowadays the revised, low
oxygen abundances measured with XMM, Mushotzky, this conference).

In brief, a wealth of work in literature suggests that some non--standard
scenario (or IMF) must be invoked to account for the metals in the ICM. 
We recall that a non--standard IMF has been suggested for elliptical 
galaxies also on the base of other, independent arguments:
\begin{itemize}
\item
a top--heavy IMF ($x \sim 1.0$) is better suited to reproduce the photometric 
properties of ellipticals (Arimoto \& Yoshii 1987);
\item
sistematic variations of the IMF in ellipticals of increasing mass might 
explain the increase of the $M/L$ ratio with galactic luminosity, that is the
so--called ``tilt of the Fundamental Plane'' (Larson 1986, Renzini \& Ciotti
1993, Zepf \& Silk 1996).
\end{itemize}
\subsection{The amount of intra-cluster gas}
Just as in the early work by Matteucci \& Vettolani (1988), most authors
conclude that GWs cannot account for the huge amount of gas present
in the ICM (2--5 times the mass in galaxies, Arnaud et~al.\ 1992). 
The ICM must then consist, for a 50 to 90\%, of primordial gas.

Trentham (1994), on the base of the steep slope of the LF at the low 
luminosity end observed in clusters, suggested that all the intra--cluster
gas could have originated in dwarf galaxies, since these are
numerous in clusters and they are expected to eject a large fraction of 
their initial mass as GW, due to their shallow potential wells.
This suggestion
was discarded by Nath \& Chiba (1995) and by Gibson \& Matteucci (1997a),
who calculated detailed models of dwarf galaxies and related GW ejection
to show that galaxies cannot be the only source for the whole intra--cluster 
gas, even in the case of the steepest observed LF
(hence the largest contribution from dwarfs).
\section{A non--standard IMF}
\label{section2}
As mentioned in \S~\ref{metalcontent}, many authors have suggested 
that some non-standard
IMF must be invoked to explain the amount of metals in the ICM. What physical
reason may justify a different IMF in different situations?

Larson (1998) suggested the following functional form of the IMF:
\[ \frac{dN}{d\log{M}} \propto M^{-x} exp \left( - \frac{M_s}{M}\right) \]
where $M_s$ is a typical mass scale related to the Jeans mass.
In brief, this IMF is a (Salpeter--like) power law down to a typical
peak mass
\[ M_p \sim \frac{M_s}{x} \]
below which there is an exponential cut-off. The peak mass
varies with the temperature and density conditions 
of the parent gas as
\[ M_p \propto T^2 \, \rho^{-\frac{1}{2}}\]
as expected from Jeans' law.
In warmer and/or less dense gas, therefore,
the typical peak mass $M_p$ increases; namely, relatively more massive
stars are formed while less mass remains locked into ever-living, very 
low--mass stars (lower locked-up fraction, e.g.\ Tinsley 1980).

A similar behaviour is predicted by the theoretical IMF by 
Padoan et~al.\ (1997, hereinafter PNJ), which features a peak mass 
\[ M_p = 0.2 M_{\odot} \left( \frac{T}{10 K} \right)^2  
\left( \frac{n}{1000 K} \right)^{-\frac{1}{2}} 
\left( \frac{\Sigma_g}{2.5~km~sec^{-1}} \right)^{-1} \]
Although the physical derivation of the PNJ IMF has been sometimes questioned 
(e.g.\ Scalo et~al.\ 1998), one can still adopt it 
in galactic models as a tentative recipe, yielding the behaviour 
na{\"\i}vely expected for the typical Jeans mass; see Chiosi (2000) for 
further discussion.

\subsection{Galactic models with the PNJ IMF}
Chiosi et~al.\ (1998) developed chemo--thermodynamical models following
the thermodynamical evolution of the gas in an elliptical galaxy,
and the corresponding variations in the IMF according to the PNJ recipe. 
The characteristics and behaviour of these models as a function of galactic 
mass  and redshift of formation are discussed in full details in 
Chiosi et~al.\ (1998) and Chiosi (2000). Here, we briefly underline 
the qualitative trends with respect to (a) mass and (b) redshift of formation.
\begin{description}
\item[(a)]
At increasing galactic mass, the average density of the object $\rho$ 
decreases and the typical peak mass $M_p$ increases, 
yielding a lower locked-up fraction.
\item[(b)]
At increasing redshift $z_{for}$ of formation, the temperature
of the object increases, since it can never fall below the corresponding
temperature of the cosmic microwave background, $T \geq T_{CMB}(z_{for})$, 
which increases with redshift; hence, the peak mass $M_p$ is higher and the 
locked--up fraction is lower.
\end{description}
We remark here that the above mentioned trends are by no means drastic:
the peak mass $M_p$ never exceeds 1~$M_{\odot}$, and the 
``high $M_p$'' phase is limited to the early galactic ages; after the initial
stage, in fact, the system reaches a sort of thermodynamical balance, 
with the peak mass and the IMF settling on quite standard values.
The overall picture loosely resembles the bimodal behaviour suggested
by Elbaz et~al.\ (1995), with an early phase dominated by massive stars
followed by a more normal star formation phase producing the low--mass
stars we see today. However, in our models
the IMF naturally and smoothly varies in time following a physical 
prescription.

Though not long-lasting, the variations in the early phases are enough to
differentiate the resulting galactic models, making them
successful at reproducing many features of observed ellipticals, 
such as (Chiosi et~al.\ 1998): 
\begin{itemize}
\item
the tilt of the Fundamental Plane (\S~\ref{metalcontent})
\item
the analogous of the ``G--dwarf'' problem, or the lack of a large population 
of low metallicity stars, detected
in the spectral energy distribution of ellipticals (Bressan et~al.\ 1994,
Worthey et~al.\ 1996);
\item
the high fraction of white dwarfs (Bica et~al.\ 1996);
\item
both the colour--magnitude relation {\it and} the trends in 
$\alpha$--enhancement with mass {\it at the same time}, thereby
overcoming the well-known dichotomy between the ``classic'' and 
``inverse'' GW scenario (Matteucci 1992, 1994, 1997).
\end{itemize}
This last point is worth commenting further, as the GW modelling influences
directly the predictions concerning the ICM. 
GW models of elliptical galaxies with a constant IMF (whether Salpeter
or more top--heavy) face the following puzzle.
The colour--magnitude relation suggests that GWs occur
{\it later} in more massive ellipticals than in smaller ones, 
so that star formation and chemical enrichment proceed longer 
and the stellar population reaches redder colours in more luminous objects. 
On the other hand, metallicity indices, if interpreted as abundance 
indicators, suggest that the [Mg/Fe] ratio increases with galactic mass; 
this requires GWs to occur {\it earlier} in more massive galaxies, where
only SN~II should contribute to the chemical enrichment to make the resulting
abundance ratios in stars $\alpha$--enhanced. 

This dichotomy between the so--called ``classic'' and ``inverse'' 
GW scenario, hampers predictions
of the metal pollution of the ICM, since two competing sets of GW models
are to be considered. It is therefore quite appealing, when we address
the chemical enrichment of the ICM, that the variable IMF scenario described
above can reproduce both observational constraints,
with a unique set of models.

\begin{figure}
\centerline{\psfig{figure=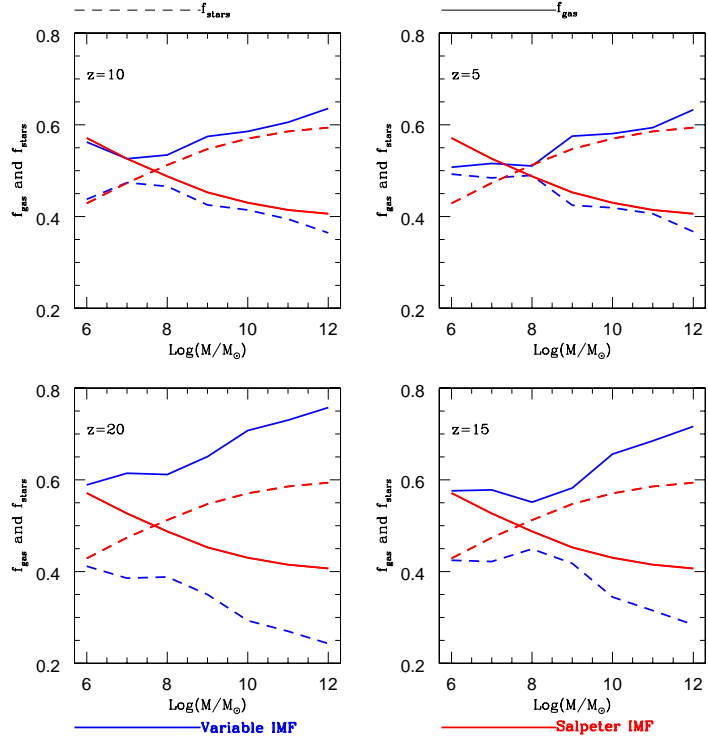,width=10truecm}}
\caption{Mass fraction of ejected gas ({\it solid lines}) and complementary
fraction locked into stars ({\it dashed lines}) as a function of the
initial (baryonic) galactic mass. The four panels correspond to models
with different redshifts of formation, as indicated.
{\it Red lines}: galaxy models with Salpeter IMF; {\it blue lines}:
models with variable PNJ IMF.}
\label{fig:ejgas}
\end{figure}

\subsection{Galactic ejecta: PNJ vs.\ Salpeter IMF}
\label{GWejecta}
Chiosi (2000) first analyzed what are the predictions about the metal 
pollution of the ICM when galaxy models with the PNJ IMF are adopted.
To this purpose, he calculated multi--zone chemical models of ellipticals 
with the PNJ IMF.

The adoption of radial multi--zone models, rather than simple
one--zone models, has in fact noticeable consequences on
the predicted enrichment of the ICM, as first underlined by
Martinelli et~al.\ (2000).
When the radial structure of an elliptical galaxy is considered,
with the corresponding gradients in density, colours etc., it turns out
that the GW tends to develop not instantly over the whole galaxy, but 
earlier in the outskirts
(where the potential well is shallower) and later in the central parts.
This means that star formation and chemical enrichment proceed longer
in the centre than in the outer regions (Tantalo et~al.\ 1998, Martinelli 
et~al.\ 1998), and the GW ejected from different galactic regions is 
metal enriched to different degrees.

The models calculated by Chiosi (2000) account for these effects
by dividing the galaxy into three zones: a central sphere where
star formation and metal production is most efficient and lasts longer;
an intermediate shell where the GW sets in earlier and the metal production
proceeds to a lesser extent, and an outer corona where the gas is expelled
almost immediately, with virtually no star formation and chemical processing.
This behaviour is the combined result of the shallower potential well when
moving outward in the galaxy (as in standard models with a constant IMF)
and of the varying $M_p$ in the PNJ IMF when moving to outer, less dense
regions; see Chiosi (2000) for a detailed discussion.

\begin{figure}[t]
\centerline{\psfig{figure=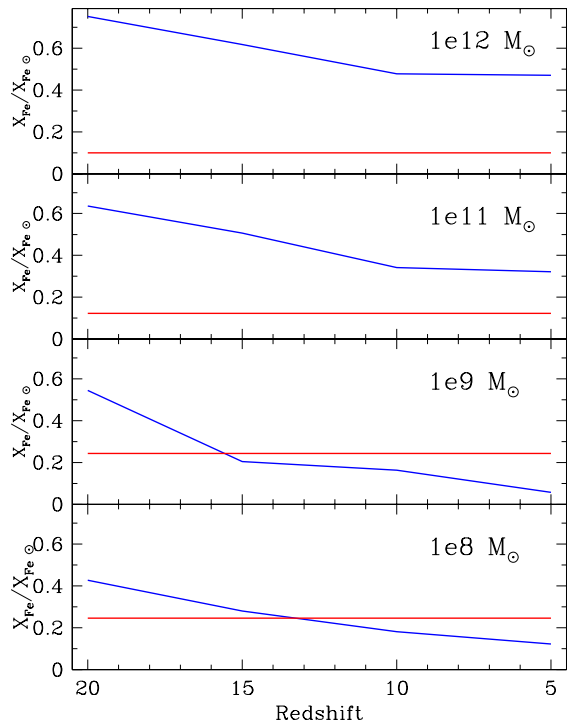,height=10.4truecm}}
\caption{Metallicity (iron abundance) of the gas ejected as GW
from galaxies of given initial (baryonic) mass, indicated 
in the individual plots, 
and as a function of their redshift of formation. {\it Blue lines}:
galactic models with the PNJ IMF; {\it red-lines}: comparative results
for models with the Salpeter IMF, which are redshift independent.
\label{fig:abundGW}}
\end{figure}

For the sake of comparison, analogous models with the standard Salpeter IMF
were also calculated.
For a better understanding of the results concerning the chemical evolution
of the ICM, let's first inspect the predicted GW ejecta of the galactic models
when the variable IMF or the Salpeter IMF are adopted in turn.

In Fig.~\ref{fig:ejgas} we compare the mass fraction of gas ejected in the GW,
and the complementary mass fraction locked into stars, for galactic models
with the variable PNJ IMF and for models with the Salpeter IMF
(blue and red lines, respectively). Mass fractions refer to
the total initial baryonic mass of the
(proto)galaxy. The amount
of ejected gas is larger in the case of the PNJ IMF, since less mass is locked
into low-mass stars, thanks to the high $M_p$ in the early
galactic phases. The difference with the Salpeter case gets sharper
for larger (proto)galactic masses, and for higher redshifts of formation.
(Notice that models with the Salpeter IMF bear no dependence on the
redshift of formation, as there are no temperature effects on the IMF
in this case).

It is worth underlining here the following ``inverse'' behaviour of the
models with the PNJ IMF with respect to what is generally expected 
from models with a constant IMF. According to the general consensus,
larger galaxies store a larger fraction of their mass into stars and
eject a lower fraction of gas in the GW, while smaller galaxies are
more efficient in wind ejection, due to their shallower potential wells.
This behaviour is in fact found in the Salpeter galactic models of
Fig.~\ref{fig:ejgas}. The models with the PNJ IMF, on the other hand, show
quite the opposite behaviour: larger galaxies eject a larger fraction of their
initial mass in the wind and lock a lower fraction into stars, due to the 
higher peak mass that characterizes them in the early phases. This effect
overwhelms their deeper potential wells, and the trend becomes stronger and
stronger with the redshift of formation.

Fig.~\ref{fig:abundGW} shows the iron abundances in the gas ejected as GW,
again comparing the Salpeter IMF (red lines) and the PNJ IMF case (blue lines).
In most cases, the galactic ejecta in the PNJ models are more metal--rich 
than in the Salpeter case, up to a factor of 5 or more in the case of the
more massive galaxies, and for high redshifts of formation. 
In the PNJ models, in fact, more gas in the galaxy gets recycled
through massive stars, effective metal contributors, and less gets locked into
low--mass stars, before the GW occurs.

\bigskip
From the trends described above,  we expect that models of ellipticals
with the PNJ IMF predict, for the ICM, a more efficient metal pollution
and a higher fraction of the gas originating from GWs,
with respect to ``standard'' models. The first results in this respect
are discussed in Chiosi (2000).

\section{The chemical evolution of the ICM: a toy model}
\label{section3}
As mentioned in \S~\ref{metalcontent}, the most popular way to calculate
the expected properties of the ICM of a cluster, on the base 
of its population of elliptical galaxies, is to make use of a grid 
of galactic models 
and integrate their GW ejecta over the observed LF, a method first developed
by Matteucci \& Vettolani (1988). In this approach, all the ellipticals
in the cluster are assumed to be coeval, of an age around 13-15~Gyrs. 

If we are to adopt here galactic models with the PNJ IMF, however,
results are expected to be very sensitive to the exact epoch (redshift)
of formation of the individual galaxies. The usual integration 
over the present--day LF is not a viable method in our scenario: 
the present-day luminosity of a galaxy no longer provides enough 
information to identify it (in terms of its initial protogalactic
mass) since the final properites of a galactic model depend not only 
on its initial mass, but also on its exact formation redshift. 
Hence, we need to follow in detail the epoch of formation 
of the individual galaxies in the cluster,
and evolve the overall system down to the present--day, using then
the LF as a constraint {\it a posteriori}. 
This approach also allows, in principle,
a more self--consistent description of the chemical evolution of the cluster
and of its galactic population as a whole.

An improved modelling of the evolution of the cluster, taking into 
account that its galaxies may form at different redshifts,
has been introduced by Chiosi (2000), who replaced the usual integration
over the LF with an integration over the Press-Schechter mass function
at different redshifts. Suitable redshift--dependent mass limits 
for the galaxies forming at each epoch were taken into account
(Tegmark et~al.\ 1997).

On the same line, we developed a global, self-consistent
chemical model for the cluster as a whole, which could follow the
simultaneous evolution of all its components: the galaxies, the primordial
gas, and the gas processed and re-ejected via GWs (Moretti et~al.\ 2001).
Our chemical model for clusters is developed in analogy with the usual
chemical models for galaxies. These latter
are schematically conceived as follows (e.g.\ Tinsley 1980, Pagel 1997):
\begin{enumerate}
\item
the gas present in the system (usually starting from primordial composition) 
keeps transforming into stars according to some prescribed Star
Formation Rate (SFR);
\item
stars are thus formed, distributed according to the adopted IMF;
\item
stars return part of their mass in the form of chemically enriched gas,
according to the so--called stellar yields (prescriptions derived from
stellar evolution and  nucleosynthesis);
\item
this chemically enriched gas mixes with the surrounding gas, causing
the chemical evolution of the overall interstellar medium (ISM).
\end{enumerate}
In a cluster, we are interested to model the chemical evolution of the
ICM, and the objects responsible for its enrichment are the galaxies,
via GWs.
In analogy with the above scheme, therefore, our chemical model 
for the cluster is conceived as follows:
\begin{enumerate}
\item
the primordial gas in the ICM gets consumed in time by galaxy formation 
according to some prescribed Galactic Formation Rate (GFR);
\item
at each time (redshift) galaxies form distributed
in mass according to a Galactic Initial Mass Function (GIMF), derived from the 
Press-Schechter mass function suited to that redshift;
\item
galaxies restitute a fraction of their initial mass in the form of
chemically enriched GWs, according to the adopted galactic models;
\item
this enriched gas mixes with and causes the chemical evolution of the 
overall ICM, which includes the amount of primordial gas not yet consumed 
by galaxy formation (if any) and the gas re-ejected by galaxies in the GWs 
up to the present age.
\end{enumerate}
We can schematically plot the analogy between the two types 
of models as follows:

\bigskip
\begin{center}
\begin{Large}
\begin{tabular}{c c c|c c c}
	    && ~~~~~\fbox{primordial gas}~~~~~	& 
~~~~~\fbox{primordial gas}~~~~~&& \\

	    && $\Downarrow$ 			& $\Downarrow$ && \\

& $\swarrow$ & {\normalsize SFR, IMF}		& {\normalsize GFR, GIMF} &
$\searrow$ & \\

	    && $\Downarrow$ 			& $\Downarrow$ && \\

ISM 	    && \fbox{stars}			& \fbox{galaxies} && ICM \\

	    && $\Downarrow$ 			& $\Downarrow$ && \\

& $\nwarrow$ & {\normalsize stellar yields} 	& {\normalsize GW yields} &
$\nearrow$ & \\

	    && $\Downarrow$ 			& $\Downarrow$ && \\

	    && \fbox{enriched gas}		& \fbox{enriched gas} && \\

\end{tabular}
\end{Large}
\end{center}

\bigskip\noindent
Model equation parallel those of galactic chemical models, with the 
substitutions {\mbox{SFR $\rightarrow$ GFR}}, 
{\mbox{IMF $\rightarrow$ Press-Schechter GIMF}}, 
{\mbox{stellar yields $\rightarrow$ GW yields}}. For further details on the
model and its equations, see Moretti et~al.\ (2001). Here we only 
point out a few main assumptions at the base of the model:
\begin{itemize}
\item
The model is one--zone, namely the cluster is treated as a single uniform
compound of gas and galaxies, where the metal abundance of the gas  evolves
in time but is otherwise homogeneous in space.
\item
The model is calculated assuming the Instantaneous Recycling Approximation
(IRA), that is assuming that galaxies eject the corresponding GWs instantly,
as soon as they are formed; this is a reasonable approximation as the
timescales for the onset of the GW are generally short, less than 1~Gyr.
The IRA could affect predictions at high redshifts, where
a time--span of a few $10^8$~yr corresponds to a sizeable gap in redshift; 
but up to redshift $z \sim 1$, where observational data on ICM abundances 
are available, the effect is minor.\\
Later on, the model might be improved in this respect, by taking into
account the actual delay between the formation of a galaxy and the time
when its GW is expelled; but as it is always preferable to start with 
the simplest possible assumptions, we adopt IRA for the time being.\\
Notice however that the galactic models for the individual ellipticals,
and the related GW yields,
are not calculated under the IRA approximation, but with a detailed
chemical network taking into account the different, finite stellar lifetimes
for the different stellar masses (Chiosi et~al.\ 1998 and references therein).
\item
We adopt the closed--box description for the chemical model of the cluster, 
as suggested by Renzini (1997) on the base of the homogeneity of 
the chemical properties of clusters:
open models with a variety of infall histories would lead to
a large scatter of features (metal abundances in the ICM, etc.), 
at odds with observations. Besides, the
closed--box assumption is the simplest case to consider,
when starting up a new model.\\
It is worth commenting further on how realistic it is to apply a closed box 
(i.e.\ constant mass) model to structures that are supposed to form by 
hierarchical accretion of subunits, according to current cosmological 
theories. We remind here that a one-zone chemical model contains 
no information on the spatial
distribution of its components. Therefore, at high redshifts our 
``model cluster'' can be considered simply as the sum of the subunits that
will later merge and form it, irrespectively of whether the cluster 
has in fact formed or not, as a bound gravitational structure. The chemical
model at high redshifts simply describes the average properties
of the sum of the parent subunits of the cluster.
\end{itemize}
As mentioned above, in this new approach the observed LF becomes a
constraint to compare the models with {\it a posteriori}. It turns out
in fact to be the main constraint to calibrate the model parameters, 
especially the GFR.
\section{The ``best case'' models}
\label{bestcase}
\begin{figure}
{\centering \leavevmode
\psfig{file=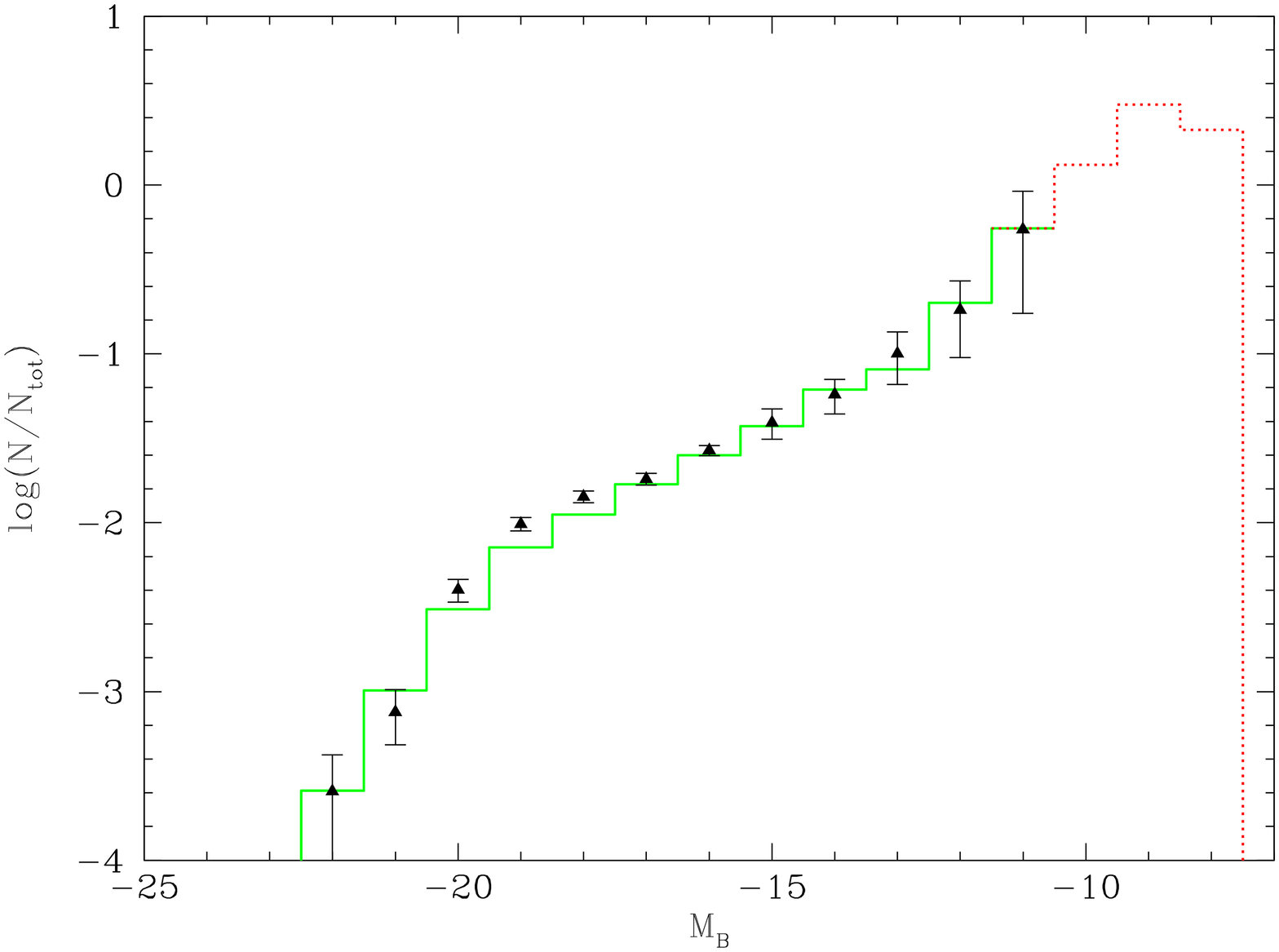,angle=-90,width=.5\textwidth} \hfil
\psfig{file=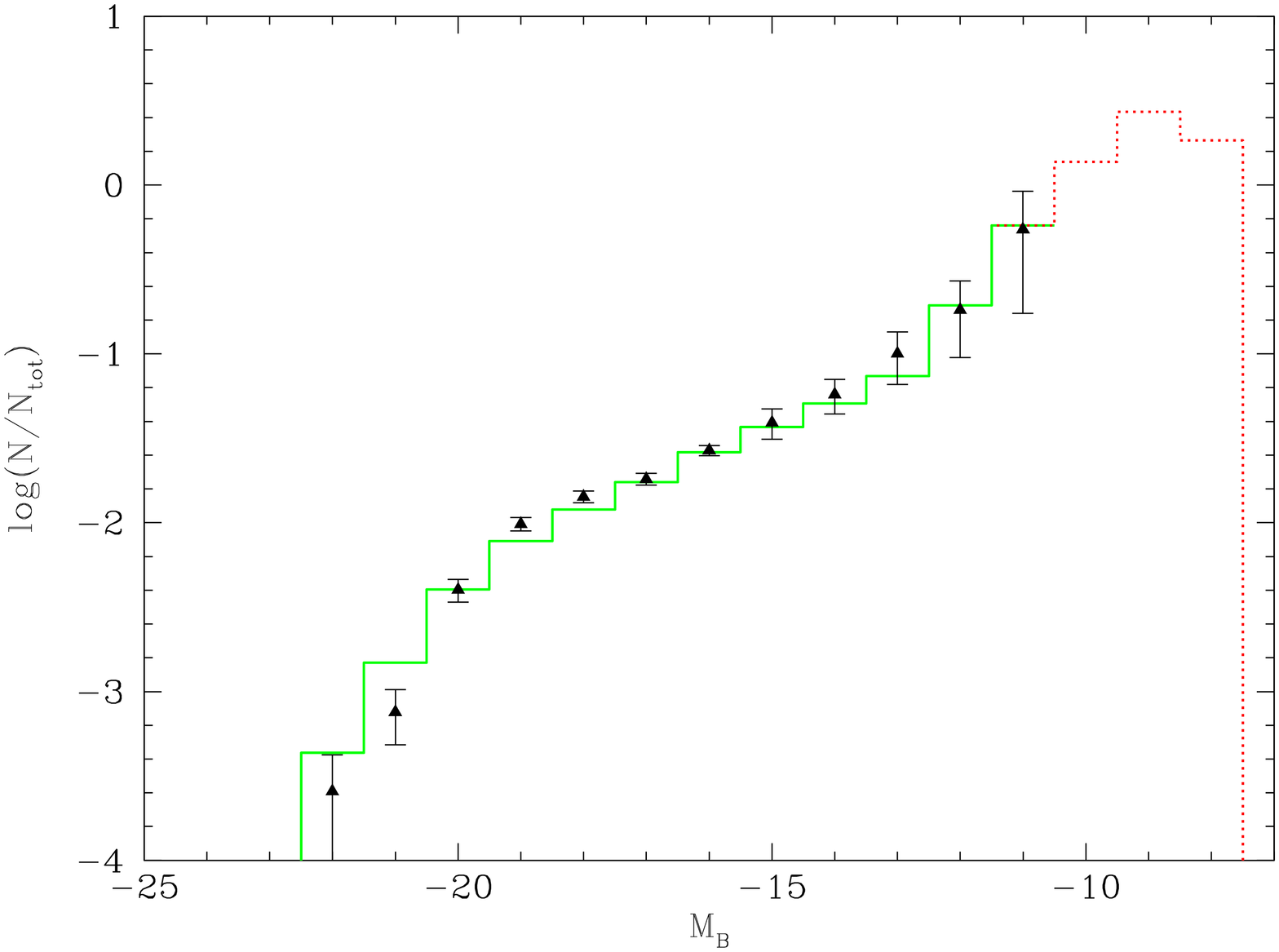,angle=-90,width=.5\textwidth}}
\caption{B--band luminosity function of cluster galaxies as predicted by our
``best case'' cluster model versus the observational
data (by Trentham 1998). {\it Left panel}: results for
galactic models with the PNJ IMF. {\it Right panel}: results for
galactic models with the Salpeter IMF. 
\label{fig:LF}}
\end{figure}

In this section we present our case of ``best match'' with the observed
LF in the B--band (Trentham 1998). This is obtained with a GFR with the
following characteristics:
\begin{itemize}
\item 
the GFR linearly increases in time from $z=20$ down to $z \sim 0.5$, 
\item
for $z < 0.5$ galaxy formation stops due to the exhaustion of the primordial 
gas out of which galaxies are assumed to form;
\item
a burst of formation of dwarf galaxies at high redshift ($z \sim 10-20$) 
is added on top of the smoothly increasing GFR to reproduce the steepening 
of the LF at low luminosities.
\end{itemize}
Here we do not address further details on the parameters of the model
and their calibration, to be found in Moretti et~al.\ (2001).
Fig.~\ref{fig:LF} shows the predicted LF compared
to the observed one, in the ``best case'' when galactic models with 
the PNJ IMF are adopted (left panel). As a comparison, the corresponding 
LF predicted with the same cluster evolution parameters, but with the Salpeter 
galactic models, is shown in the right panel. For otherwise equal 
parameters, the Salpeter case predicts more galaxies in the high--luminosity 
bins, due to the fact that for massive (proto)galaxies more mass remains
locked into stars in the Salpeter case than with the PNJ IMF 
(cf.\ Fig.~\ref{fig:ejgas}). Anyways, the LF is still in agreement
with the observed one within errors.
The dotted lines represent
the extension of the LF to the range of galaxies fainter than the
observational limit. Although these galaxies largely dominate in number,
their contribution in terms of luminosity or stellar mass is a negligible
fraction of the total; in the cluster these objects might also have been 
disrupted and be nowadays dispersed as a diffuse intra--cluster stellar 
component.

\begin{figure}
{\centering \leavevmode
\psfig{file=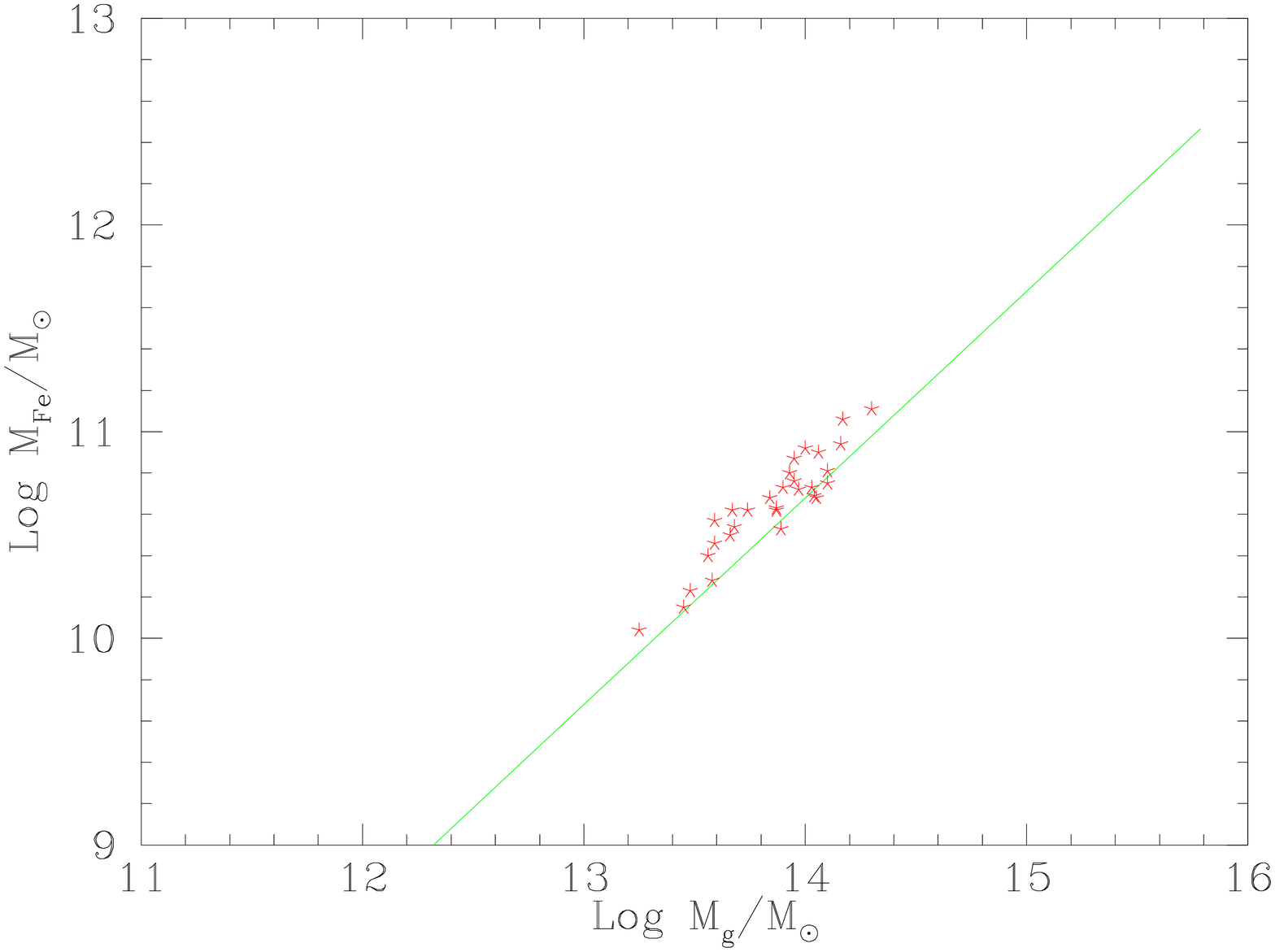,angle=-90,width=.5\textwidth} \hfil
\psfig{file=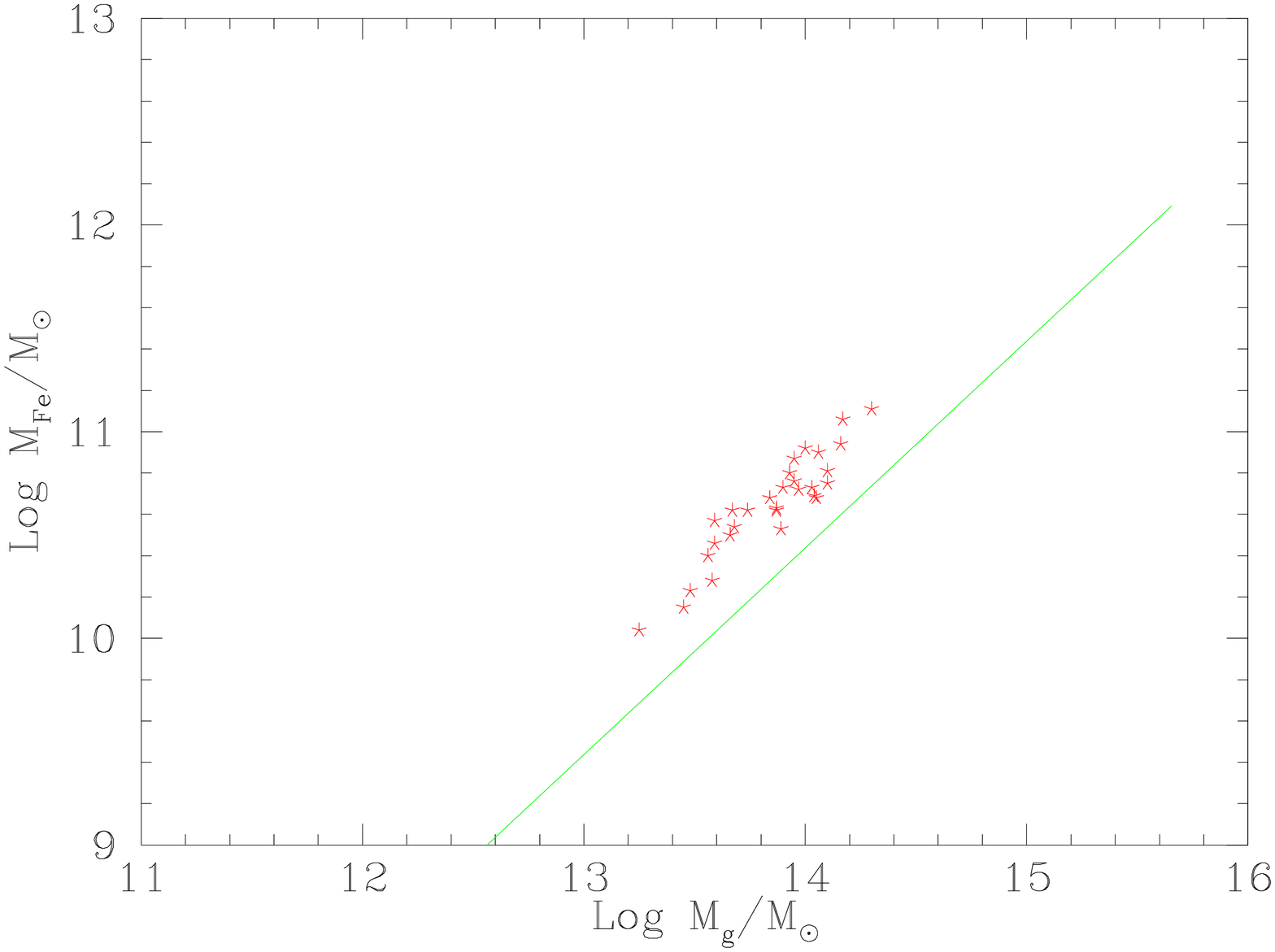,angle=-90,width=.5\textwidth}}
\caption{Metallicity (iron abundance) of the ICM as predicted 
by our chemical evolution model for the cluster versus the observational
data (by Matsumoto et~al.\ 2000). {\it Left panel}: results for
galactic models with the PNJ IMF. {\it Right panel}: results for
galactic models with the Salpeter IMF.
\label{fig:MFe}}
\end{figure}

\begin{figure}
{\centering \leavevmode
\psfig{file=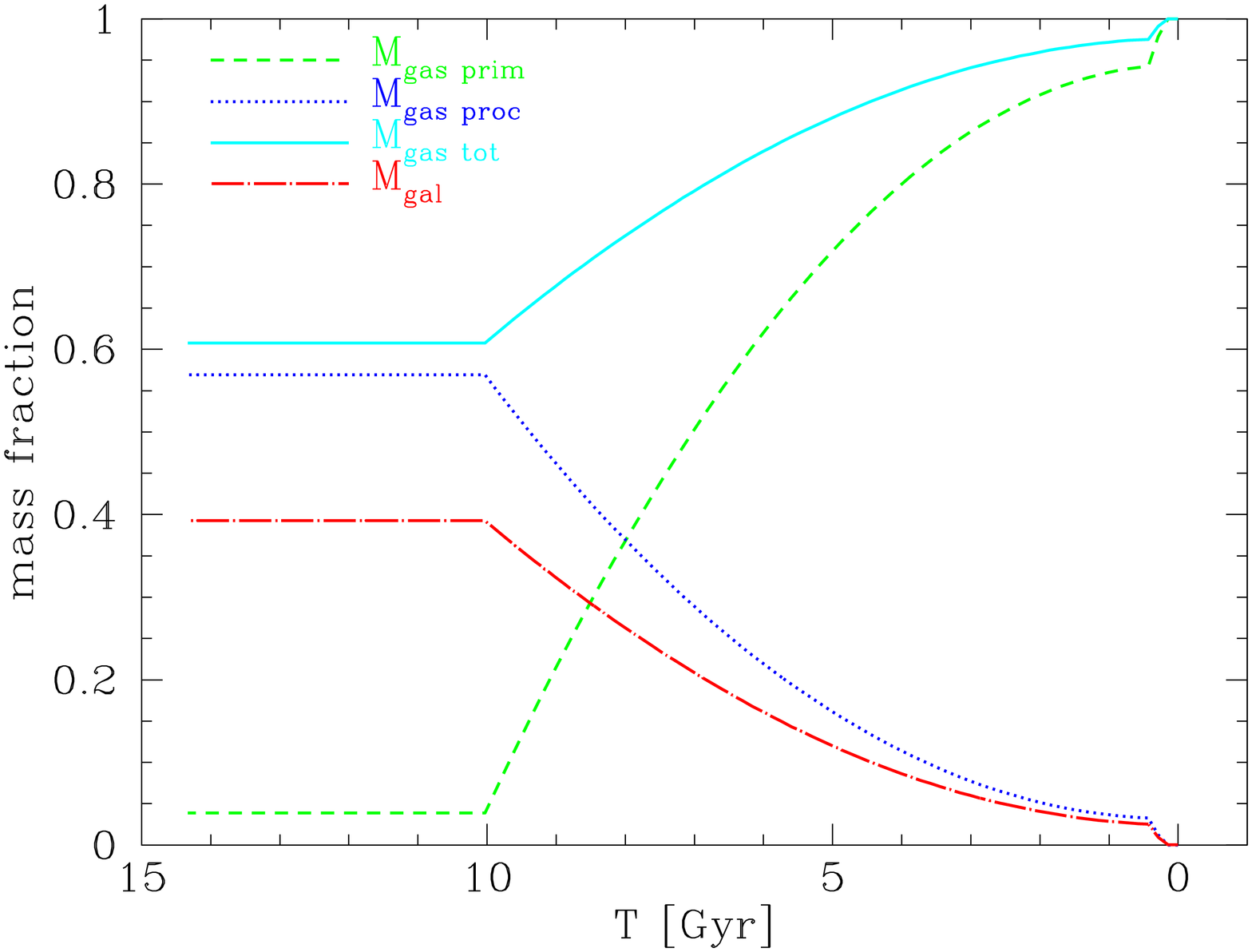,angle=-90,width=.5\textwidth} \hfil
\psfig{file=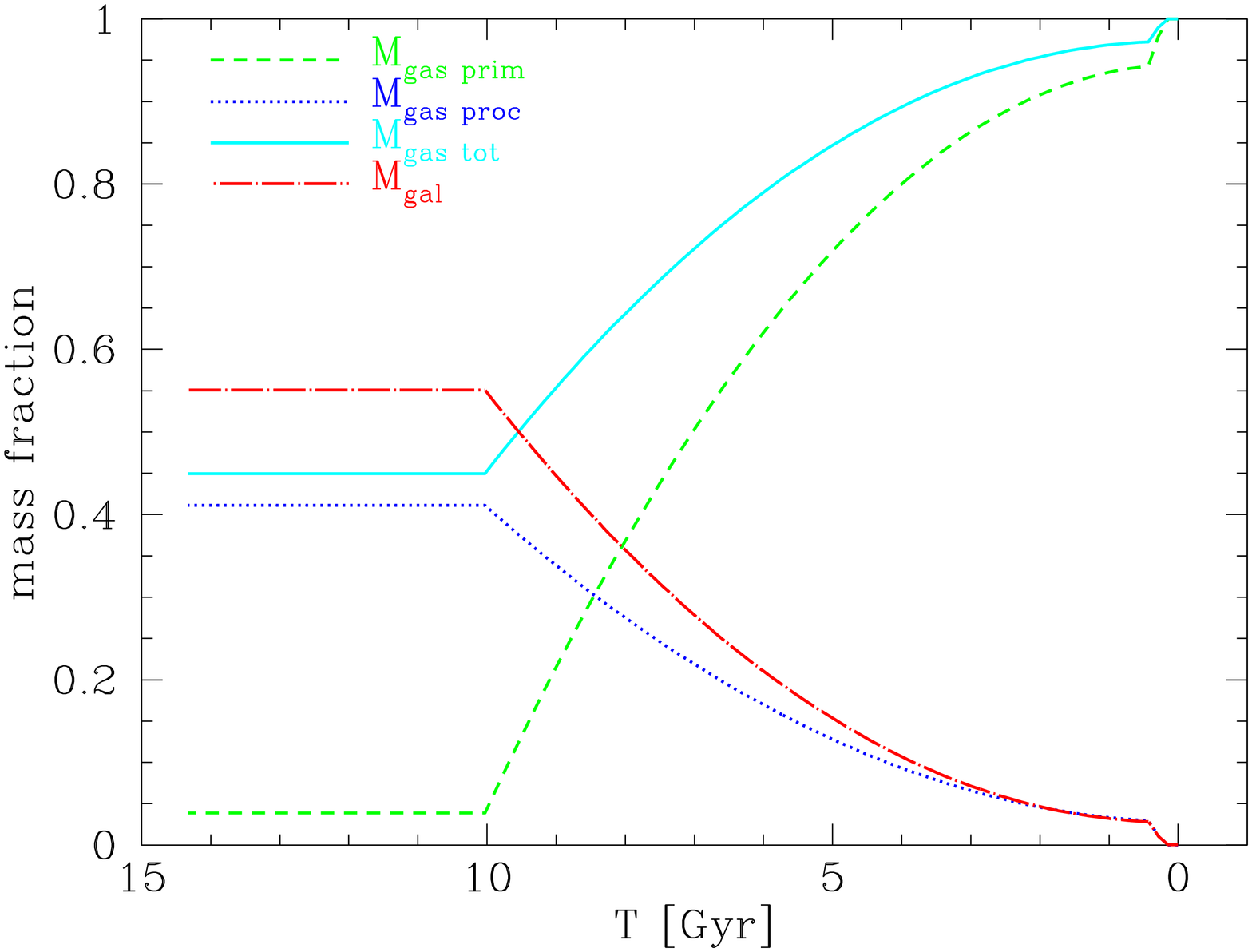,angle=-90,width=.5\textwidth}}
\caption{Time evolution of the cluster components: mass in primordial
and processed gas separately, total gas mass, and mass in galaxies.
{\it Left panel}: results for galactic models with the PNJ IMF. 
{\it Right panel}: results for galactic models with the Salpeter IMF.
\label{fig:Mgas-Mgal}}
\end{figure}

Although the predicted LF is virtually the same in the two models,
strong differences are found in the predicted gas and metallicity content 
in the ICM. Fig.~\ref{fig:MFe} shows the predicted abundances 
in the ICM in the PNJ and Salpeter case (left and right panel, respectively).
In the plot, lines of unit slope represent iso-metallicity loci. 
Observational data roughly fall onto
the same abundance level, around 0.3~$Z_{\odot}$. It is evident how
adopting galactic models with the PNJ IMF improves predictions about 
the metallicity of the ICM.

Fig.~\ref{fig:Mgas-Mgal} shows the evolution of the mass fraction
of the various components of the cluster: the primordial gas, 
which gets consumed by galaxy formation; the processed gas, 
namely the gas that has been processed inside galaxies
and then re-ejected as GW; the total gas, sum of the primordial 
and of the processed gas; the mass in galaxies, that is in the stellar
component we see today, ``left over'' after the GW. While in the Salpeter
case (right panel) the overall mass that remains locked into galaxies 
(red line) is larger that the mass ejected in the GWs (dark blue line), 
the opposite is true in the cluster model 
with the PNJ galaxies (left panel), as qualitatively expected from the 
comparison of the different galactic models in \S~\ref{GWejecta}. 
In the latter case, the mass of the re-ejected
gas is $\sim 1.5$ times larger than that locked into galaxies. Although
this is not enough to account for the whole of the observed intra-cluster gas 
(whith a mass 2--5 times larger than that in galaxies, Arnaud et~al.\ 1992), 
the amount of gas re-ejected by galaxies is expected
to make up for a remarkable fraction of the overall ICM.

\begin{figure}[hb]
\centerline{\psfig{figure=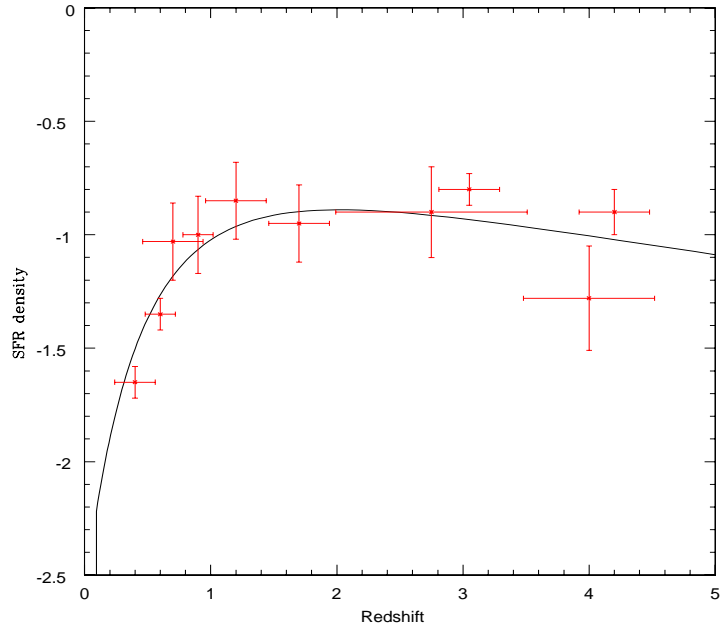,width=10truecm,height=9truecm}}
\caption{Galactic Formation Rate with a functional form following
the Madau--plot. Data Steidel et~al.\ (1999).
\label{fig:Madauplot}}
\end{figure}


\section{A Madau--like GFR}
\label{madau}
In the previous ``best case'' model the GFR was calibrated so as to
reproduce the observed LF, being otherwise arbitrary.
An alternative recipe for the GFR is a functional form resembling
the cosmic SFR of the Madau--plot 
(Fig.~\ref{fig:Madauplot}). 
The burst of dwarf galaxy formation at $z \sim 10-20$ 
is always needed to reproduce the faint end of the LF, but this is not
in constrast with the Madau--plot. 

\begin{figure}
{\centering \leavevmode
\psfig{file=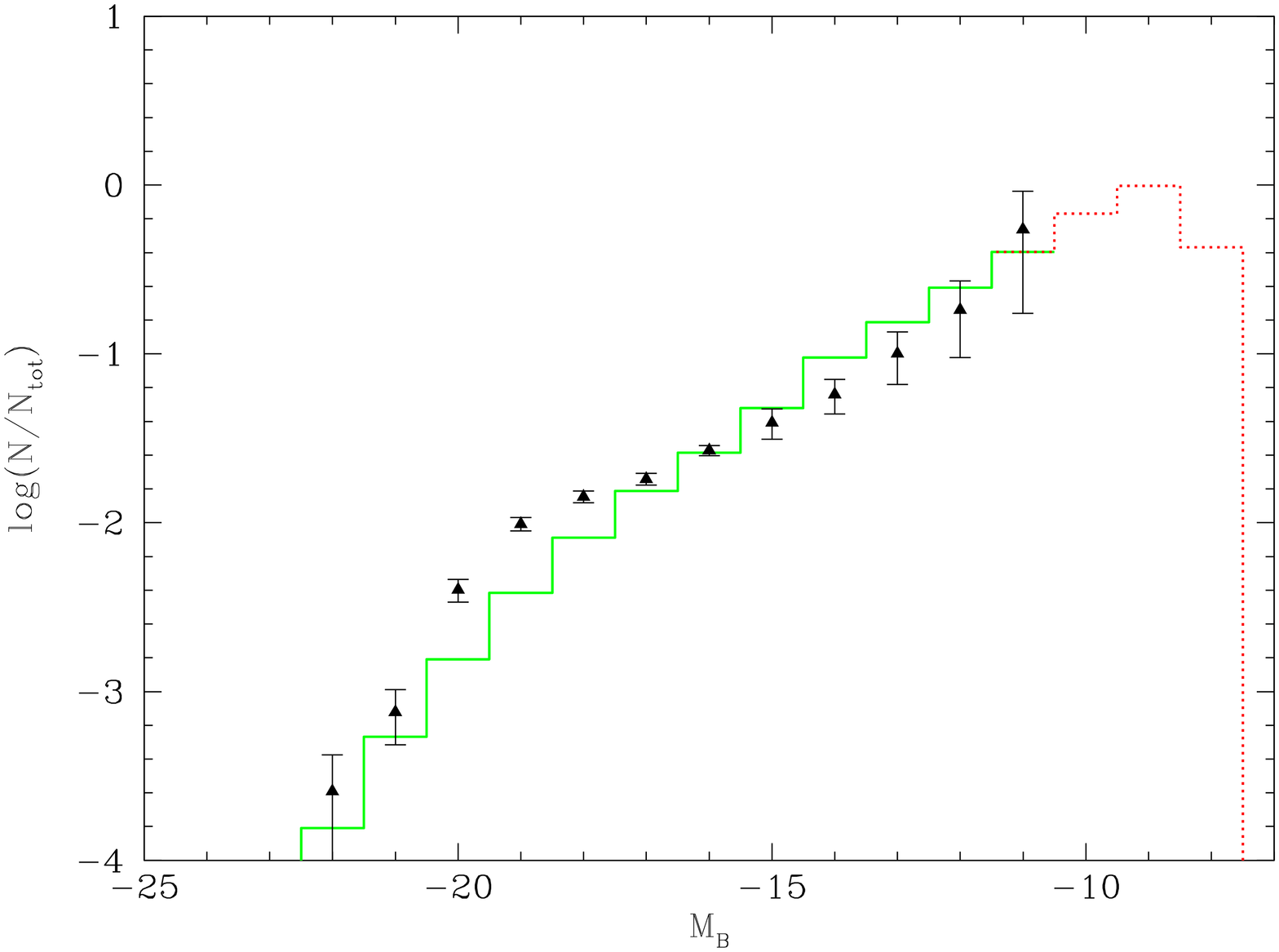,angle=-90,width=.5\textwidth} \hfil
\psfig{file=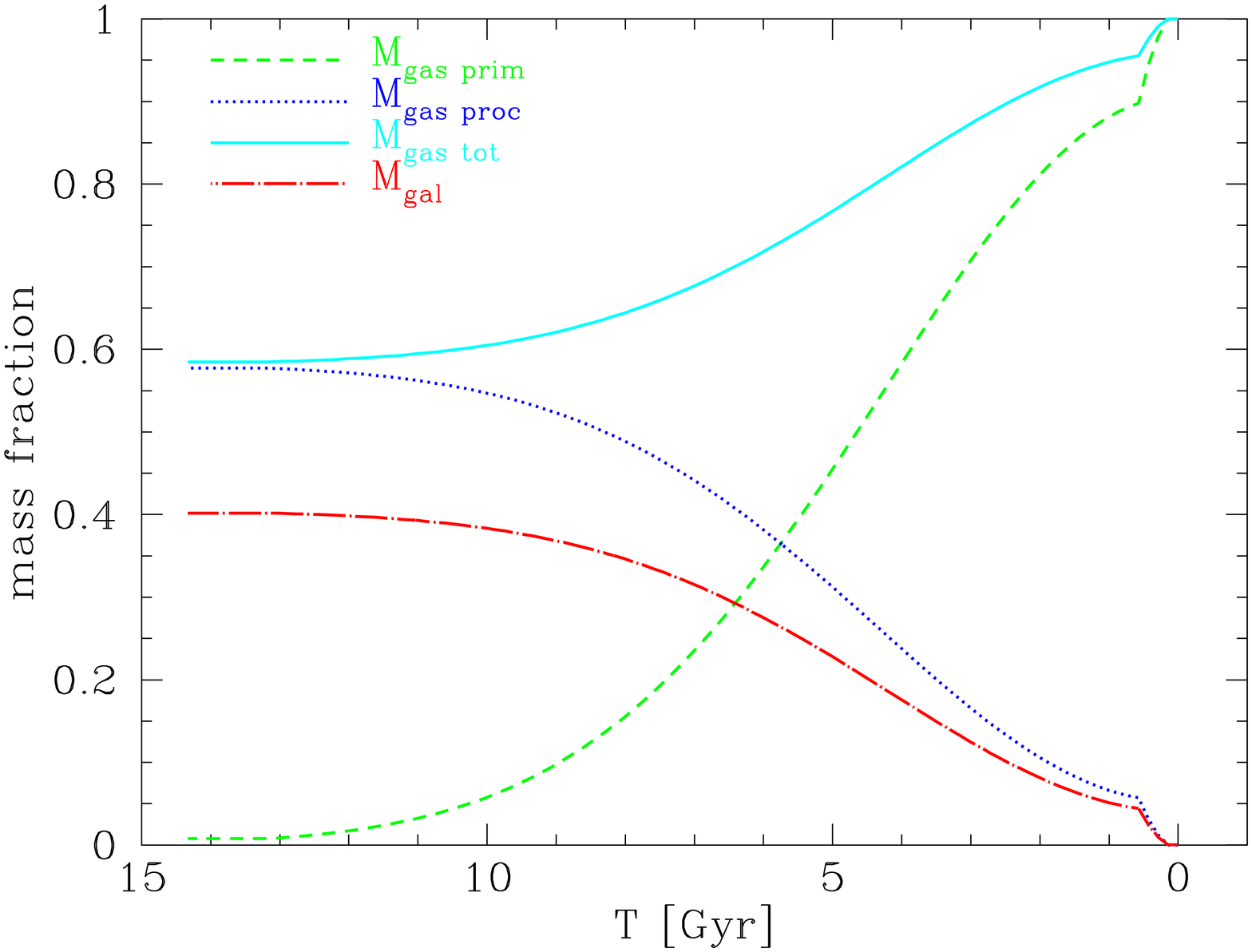,angle=-90,width=.5\textwidth}}
{\centering \leavevmode
\psfig{file=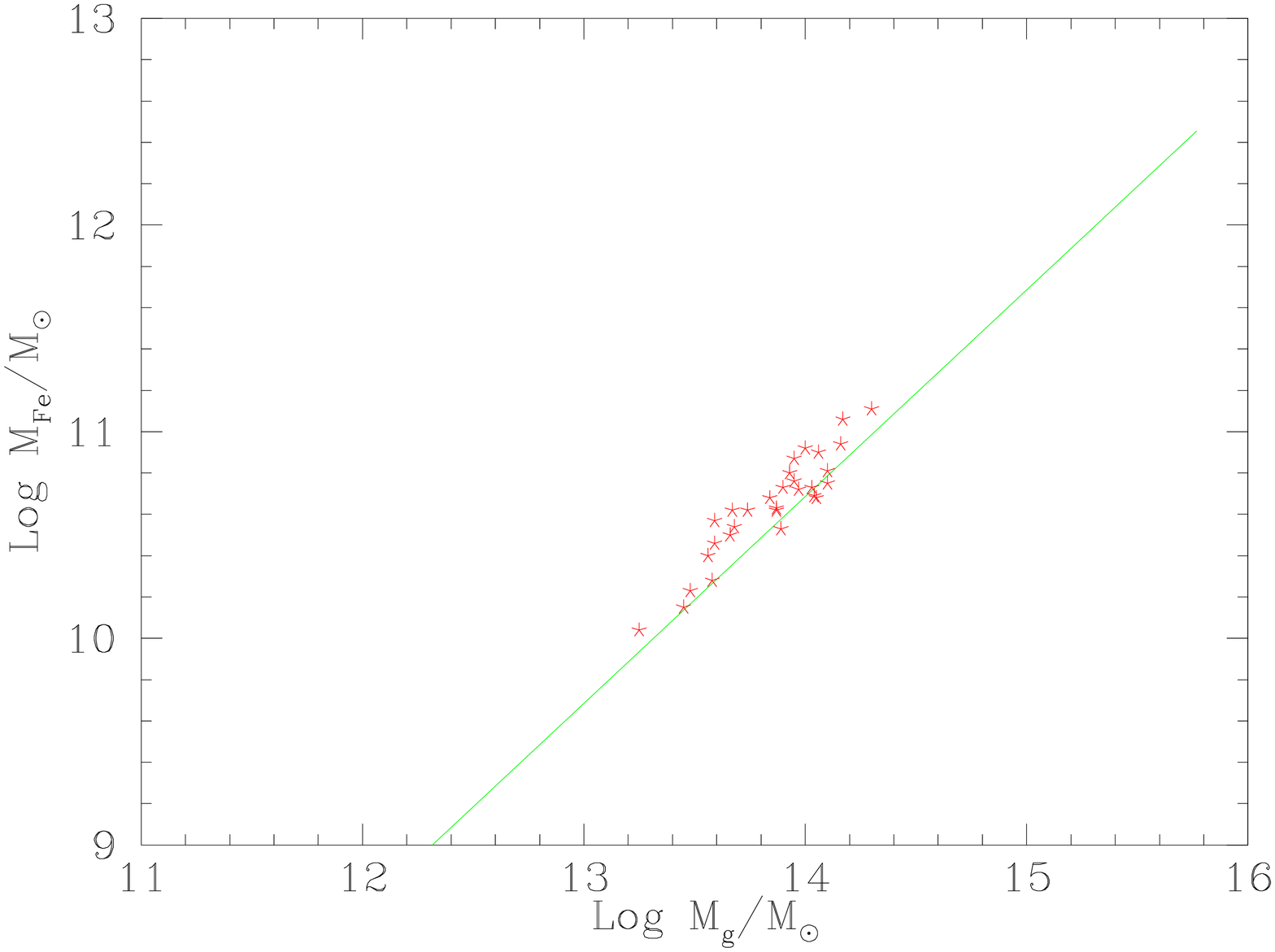,angle=-90,width=.5\textwidth} \hfil
\psfig{file=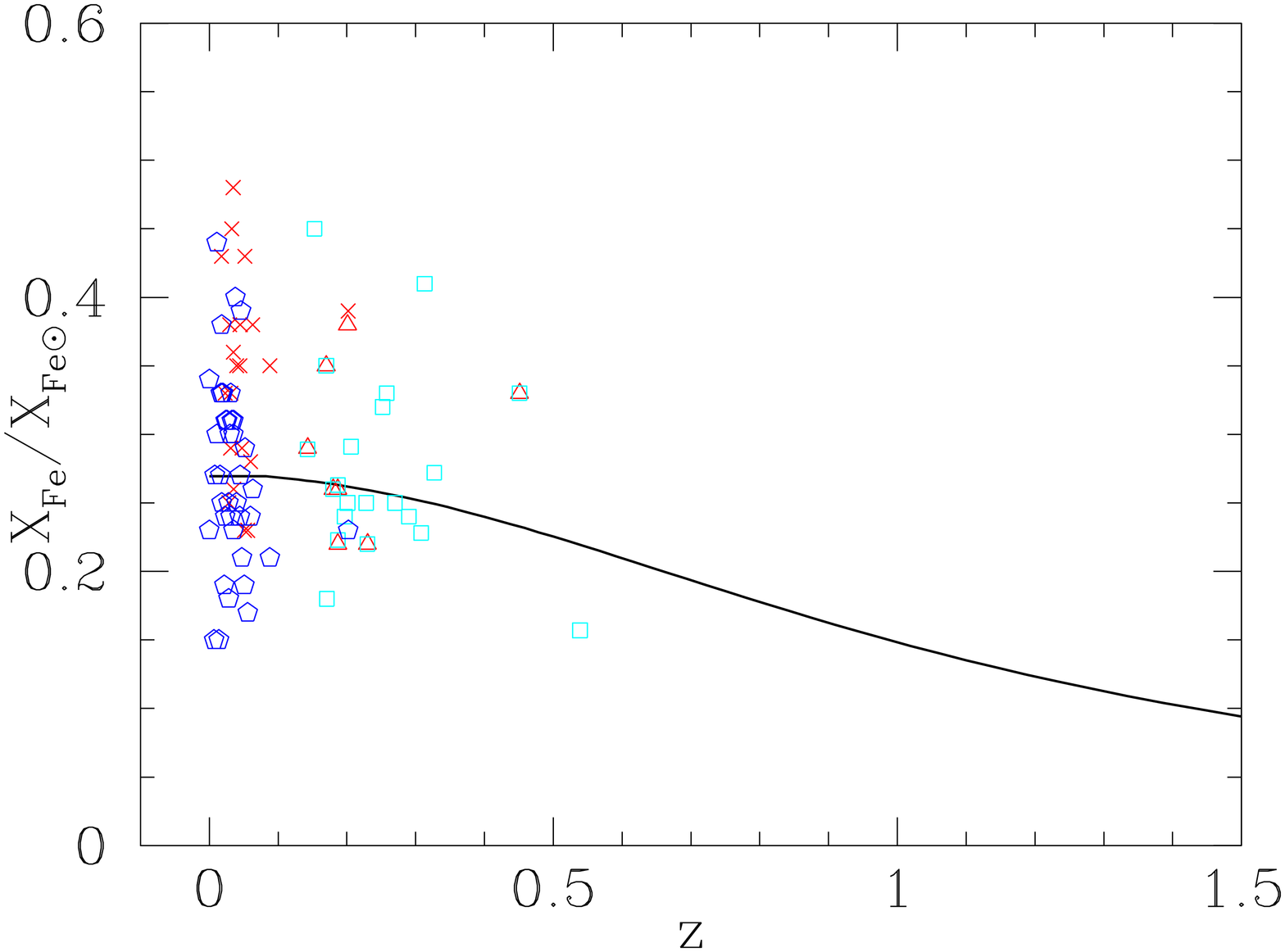,angle=-90,width=.5\textwidth}}
\caption{Results for the cluster model adopting a Madau--like GFR 
and galactic models with the PNJ IMF.
{\it Top left}: luminosity function. 
{\it Top right}: time evolution of the mass fraction in the different 
cluster components.
{\it Bottom left}: present day iron abundance in the ICM.
{\it Bottom right}: redshift evolution of the iron abundance in the ICM;
data by Matsumoto et~al.\ (2000), Fukazawa et~al.\ (1998), Mushotzky \&
Loewenstein (1997).
\label{fig:Madaures}}
\end{figure}

Fig.~\ref{fig:Madaures}
shows the corresponding predictions for metallicity, metallicity evolution
and mass fraction evolution of the various components. Only results 
for galactic models
with the PNJ IMF are shown; the comparison with the corresponding
Salpeter case would lead to considerations similar to those in 
\S~\ref{bestcase}.
Once more, the average metallicity of the observed clusters is quite well
reproduced, and a large amount of gas ($\sim$1.5 times the mass in galaxies)
is predicted to be re-ejected by GWs.
With this Madau--like form of the GFR, however, it is hard to reproduce 
the observed LF in detail.

\section{Summary and conclusions}
Galactic winds from elliptical galaxies are the most likely source of the
chemical enrichment of the ICM. In this scenario, various studies 
in literature suggests that a non--standard IMF must be invoked 
for elliptical galaxies, if we are
to account for the metal pollution of the ICM (\S~\ref{section1}).
 
We considered
GW models for ellipticals adopting a variable, physically grounded
IMF whose behaviour in time and space follows common expectations for
the Jeans mass. This IMF (by PNJ) naturally predicts a lower locked--up 
fraction 
in the early galactic stages, especially for massive ellipticals and/or 
for high redshifts of formation. Galactic models calculated with this IMF
have been shown to reproduce successfully a variety of 
observational properties of ellipticals (Chiosi et~al.\ 1998).
In this paper we discussed the resulting GW ejecta for these model ellipticals
with respect to models adopting the standard Salpeter IMF.
Our new models predict galaxies to eject both {\it more metals} and 
{\it more gas} than standard chemical models do (\S~\ref{section2}).

Then we have developed a toy--model for the chemical evolution of the ICM,
following the formation history of cluster galaxies and the corresponding
chemical evolution of the ICM in a self--consistent fashion. The model
can adopt different sets of galactic models in turn, so as to explore the
corresponding variation of the predicted metal pollution history of the ICM
(\S~\ref{section3}).

The galaxy formation history can be calibrated so as to reproduce the observed
present--day LF of cluster galaxies. A good match with the LF can be
obtained both using model ellipticals with the PNJ IMF, and using more
standard galactic models with the Salpeter IMF. However, when the PNJ
galactic models (also favoured on the base of their
spectro--photometric properties) are adopted, predictions on the metal
abundances of the ICM are remarkably improved. With these models, 
the mass of gas re-ejected in the GWs exceeds the mass stored in stars 
by a factor
of 1.5, and becomes an important fraction of the total intra-cluster gas
(\S~\ref{bestcase}).

We also explored an alternative cluster model, where the galactic formation
history follows the trend suggested by the Madau plot (at the expence of
a less successful prediction on the LF). Conclusions about the chemical
evolution of the ICM, with the PNJ IMF, do not substantially change: 
the model predicts the
correct metal abundance of the ICM, and suggests that an important fraction
of the intracluster gas may be of galactic origin (\S~\ref{madau}).

\section*{Acknowledgments}
LP acknowledges financial support from a EU grant for young European 
scientists to attend the conference.

\end{document}